\documentclass[1p,12 pt, preprint,review]{article}
\usepackage{epsfig}
\usepackage{graphicx}
\usepackage{setspace}
\date{}
\def\b{\begin{equation}}
\def\e{\end{equation}}
\def\bee{\begin{enumerate}}
\def\eee{\end{enumerate}}
\begin{document}
\title{\bf Analysis of the logarithmic slope of $F_{2}$ from the Regge gluon density behavior at small $x$}

\author{G.R.Boroun $^{1}$\thanks {Email:
grboroun@gmail.com}} \maketitle {\it \centerline{ \emph{
$^{1}$Physics Department, Razi University, Kermanshah 67149, Iran
}}
\begin{abstract}
\emph{{We study of the accuracy of the Regge behavior of the gluon
distribution function for obtain an approximation relation, which
is frequently used to extract the logarithmic slopes of the
structure function from the gluon distribution at small $x$. We
show that the Regge behavior analysis results are comparable with
HERA data and also are better than other methods that expand of
the gluon density at distinct points of expansion. Also we show
that for $Q^{2}=22.4 GeV^{2}$, the $x$ dependence of the data is
well described by gluon shadowing corrections to GLR-MQ equation.
The resulting analytic expression allow us to predict the
logarithmic derivative
$\frac{{\partial}F_{2}(x,Q^{2})}{{\partial}lnQ^{2}}$ and to
compare the results with H1 data and a QCD analysis fit with MRST
parametrization input.}}
\end{abstract}
 \vspace{0.5cm} {\it \emph{Keywords}}: \emph{shadowing
correction,Regge behavior; GLR-MQ equation; Small-$x$ }\\
 \vspace{0.5cm} {\it \emph{PACS}}: \emph{
 13.85Hd, 12.38.Bx, 13.60.Hb}

\newpage
\emph{Previously several methods of relation between the $F_{2}$
scaling violations and the gluon density at low $x$ has been
suggested [1-3]. All methods are based on an approximate relation,
as using the fact that quark densities can be neglected and that
the nonsinglet contribution $F_{2}^{Ns}$ can be ignored safely. To
investigate, we have used the DGLAP evolution equations [4] for
four flavours:} \b
\frac{dF_{2}}{dlnQ^{2}}=\frac{5\alpha_{s}}{9\pi}\int_{x}^{1}{dz}G(\frac{x}{z},Q^{2})P_{qg}(z),\e
\emph{where $P_{qg}(z)=(1-z)^2+z^2$}.\\
\emph{In LO (leading order), Pretz$^{,}$s [1] shows an
approximation relation between the gluon density at the point $2x$
and the logarithmic slopes $F_{2}$ at the point $x$, as the final
relation was found:}\b
\frac{dF_{2}}{dlnQ^{2}}=\frac{5\alpha_{s}}{9\pi}\frac{2}{3}G(2x,Q^{2}).\e
 \emph{Bora$^{,}$s [2] shows similar relation based on expansion
 of the gluon distribution around $z=0$, as was found:\b
\frac{dF_{2}}{dlnQ^{2}}=\frac{5\alpha_{s}}{9\pi}\frac{3}{4}G(\frac{4}{3}x,Q^{2}).\e}
\emph{Also Gay Ducati and Concalves [3] show this expansion at an
arbitrary point of expansion. As in the limit $x{\rightarrow}0$,
the equation becomes:\b
\frac{dF_{2}}{dlnQ^{2}}=\frac{5\alpha_{s}}{9\pi}\frac{2}{3}G(\frac{x}{1-a}(\frac{3}{2}-a),Q^{2}).\e
They could conclude that the better choice is at $a=0.75$, as:\b
\frac{dF_{2}}{dlnQ^{2}}=\frac{5\alpha_{s}}{9\pi}\frac{2}{3}G(3x,Q^{2}).\e}
\emph{All relations (2,3 and 5) estimate the logarithmic slopes
$F_{2}$ with respect to the gluon distribution function at the
points $2x$, $\frac{4}{3}x$ and $3x$. In the present letter, we
extend the method using the Regge technique. We first introduce
the Regge behavior of the gluon distribution, as can be expresed
by}:\b G(x,t)=A_{g}x^{-\lambda_{g}(t)},\e \emph{ where $A_{g}$ is
a constant and $\lambda_{g}$ is the intercept
($t=ln\frac{Q^{2}}{\Lambda^{2}})$. Using this behavior and after
integrating and some rearranging, we find an approximation
relation between the $dF_{2}(x,Q^{2})/dlnQ^{2}$ and $G(x,Q^{2})$
at the same point $x$, as we have:}\b
\frac{dF_{2}}{dlnQ^{2}}=\frac{5\alpha_{s}}{9\pi}T(\lambda_{g})G(x,Q^{2}).\e
\emph{where $T(\lambda_{g})=\int_{x}^{1}dz
z^{\lambda_{g}}(1-2z+2z^2)$.\\
Relation (7) [5] helps to estimate the logarithmic slopes $F_{2}$
in the leading logarithmic approximation (LLA).} \emph{We note
also that if we wish to evolve shadowing corrections to the gluon
density, we can simply show these recombinations with respect to
Gribov, Levin, Ryskin, Mueller and Qiu (GLRMQ)[6,7] equations.
These nonlinear terms reduce the growth of the gluon distribution
in this kinematic region where $\alpha_{s}$ is still small but the
density of partons becomes so large. According to the fusion of
two gluon corrections, the evolution of the shadowing structure
function with respect to $lnQ^{2}$ corresponds with the modified
DGLAP evolution equation. So we have }\b
\frac{{\partial}F^{s}_{2}(x,Q^{2})}{{\partial}lnQ^{2}}=\frac{5\alpha_{s}}{9\pi}T(\lambda_{g})G^{s}-\frac{5}{18}\frac{27\alpha_{s}^{2}}{160R^{2}Q^{2}}
[G^{s}]^{2},\e \emph{where $R$ is the size of the target which the
gluons populate . The value of $R$ depends on how the gluon
ladders couple to the proton, or on how the gluons are distributed
within the proton. $R$ will be of the order of the proton radius
$(R\simeq5\hspace{0.1cm} GeV^{-1})$ if the gluons are spread
throughout the entire nucleon, or much smaller
$(R\simeq2\hspace{0.1cm} GeV^{-1})$ if gluons are concentrated in
hot- spot [9] within the proton.}
 \emph{We show a plot of $\frac{{\partial}F_{2}(x,Q^{2})}{{\partial}lnQ^{2}}$ in Fig.1 for a set of values of $x$ at
  $Q^{2}$
 constant at hot spot point $R=2 GeV^{-1}$, compared to the values measured by the H1 collaboration
 [10] and a fit to ZEUS data inspired by the Froissart bound[11] based on MRST input parametrization [12]. In Fig.1 we show our results of
$\frac{dF_{2}}{dlnQ^{2}}$ obtained from the Regge behavior of the
gluon density that compared with other models based on the
expansion of the gluon density. For these results, the input gluon
used from MRST parameterizations. It is clear that our results
based on this behavior are lowest that other models. Also, from
this figure one
 can see that GLR-MQ equation tamed behavior with respect to gluon
 saturation as $x$ decreases. This shadowing
 correction suppress the rate of growth in comparison with the
 DGLAP approach.
 }
 \emph{To conclude, At high density the recombination
of gluons becomes dominant, and must be included in the
calculations. When the shadowing term is combined with DGLAP
evolution in the double leading log approximation (DLLA), then we
obtain the GLR-MQ equation for the integrated gluon. So that, we
have solved the DGLAP equation, with the nonlinear shadowing term
included, in order to determine the very small $x$ behavior of the
gluon distribution $G(x,Q^{2})$ of the proton. In this way we were
able to study the interplay of the singular behavior, generated by
the linear term of the equation, with the taming of this behavior
by the nonlinear shadowing term. With decreasing $x$, we find that
an ${\sim}x^{-\lambda_{g}}$ behavior of the gluon function emerges
from the GLR-MQ equation. Based on our present calculations we
conclude that the behavior of
$\frac{{\partial}F_{2}(x,Q^{2})}{{\partial}lnQ^{2}}$ as measured
by HERA, tamed based on gluon saturation at low $x$. Our results
show that the data can be described in PQCD taking into account
shadowing corrections.}

\emph{
}

\newpage
\subsection*{Figure captions }
{\emph{Fig 1: A plot of the derivative of the structure function
with respect to $lnQ^{2}$ vs. $x$ for $Q^{2}=22.4 GeV^{2}$ with
MRST parametrization [12] , compared to data from H1 Collab.[10]
(circles) with total error, and also a QCD fit [11] and other
models [1-3](Dot curves). Solid curves are our results with and
without shadowing correction with respect to the Regge behavior of
the gluon density.}
\begin{figure}
\centering
  \includegraphics[width=15cm]{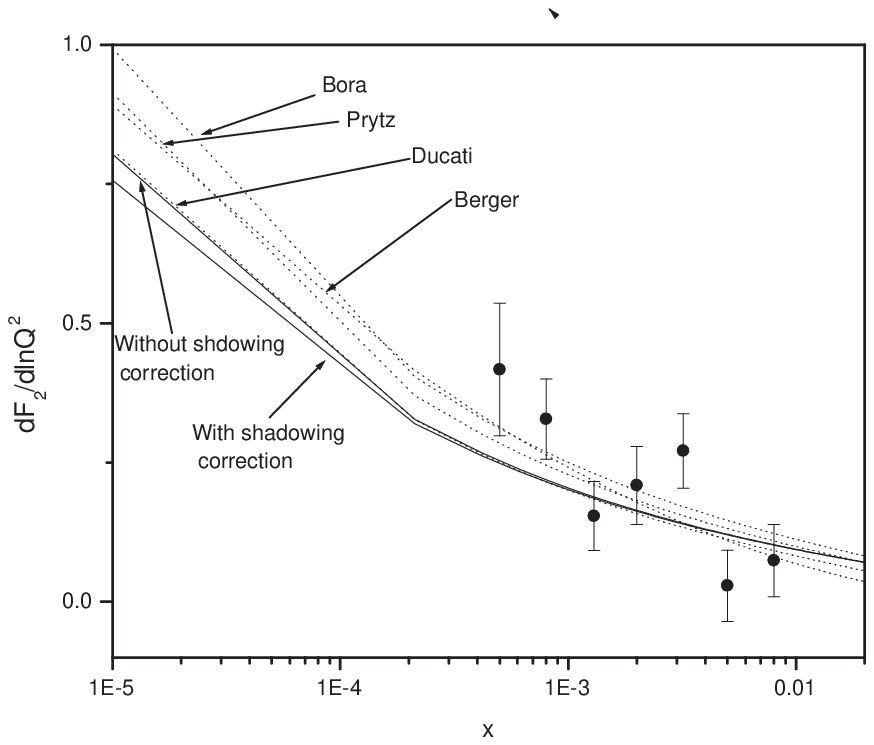}\\
  \caption{  }\label{}
\end{figure}


\begin{thebibliography}{a}
\bibitem{h1}. K.Prytz, Phys.Lett.B \textbf{311}, 286(1993); Phys.Lett.B \textbf{332}, 393(1994).
\bibitem{h1}. K.Bora and D.K.Choudhury, Phys.Lett.B \textbf{354}, 151(1995).
\bibitem{h1}. M.B.Gay Ducati and P.B.Goncalves, Phys.Lett.B \textbf{390}, 401(1997).
\bibitem{h4}. Yu.L.Dokshitzer, Sov.Phys.JETP {\textbf{46}}, 641(1977);
G.Altarelli and G.Parisi, Nucl.Phys.B \textbf{126}, 298(1977);
V.N.Gribov and L.N.Lipatov, Sov.J.Nucl.Phys. \textbf{15},
438(1972).
 \bibitem{h5}. A.V.Kotikov, JETP Lett.\textbf{59}, 667(1994);
 A.V.Kotikov and G.Parente. Phys.Lett.B\textbf{379}, 195(1996).
 \bibitem{h6}. L.V.Gribov, E.M.Levin and M.G.Ryskin, Phys.Rep.\textbf{100},
 1(1983).
 \bibitem{h7}. A.H.Mueller and J.Qiu, Nucl.Phys.B\textbf{268}, 427(1986).
 \bibitem{h8}. G.R.Boroun, JETP,\textbf{133}, No.4, 805(2008); Eur.Phys.J.A\textbf{42}, 251 (2009).
 \bibitem{h9}. E.M.Levin and M.G.Ryskin, Phys.Rep.\textbf{189}, 267(1990).
 \bibitem{h10}. $H1$ Collab., C.Adloff \textit{et al}., phys.Lett.B \textbf{520}, 183(2001); Eur.Phys.J.C \textbf{13},
609(2000); Eur.Phys.J.C \textbf{21}, 33(2001).
\bibitem{h11}. E.L.Berger, M.M.Block and C.I Tan, Phys.Rev.Lett\textbf{98},
242001(2007).
\bibitem{h12}. A.D.Martin, W.J.Stirling, R.G.Roberts,  and R.S.Thorne,
Phys.Rev.D \textbf{47}, 867(1993); J.Kwiecinski, A.D.Martin and
P.J.Sutton, Phys.Rev.D \textbf{44}, 2640(1991); A.J.Askew,
J.Kwiecinski, A.D.Martin and P.J.Sutton, Phys.Rev.D \textbf{47},
3775(1993).
\bibitem{h13}. V.Chekelian, Nucl.Phys.A\textbf{755},
111(2005).
\end{thebibliography}
\end{document}